%% file: paper.tex
\documentclass{article}
\usepackage{graphics}

\begin{document}

\title{A Methodology for Determining Amino-Acid Substitution Matrices from Set
Covers}

\author{Alexandre~H.~L.~Porto\\
Valmir~C.~Barbosa\thanks{Corresponding author ({\tt valmir@cos.ufrj.br}).}\\
\\
Universidade Federal do Rio de Janeiro\\
Programa de Engenharia de Sistemas e Computa\c c\~ao, COPPE\\
Caixa Postal 68511\\
21941-972 Rio de Janeiro - RJ, Brazil}

\date{}

\maketitle

\begin{abstract}
We introduce a new methodology for the determination of amino-acid substitution
matrices for use in the alignment of proteins. The new methodology is based on
a pre-existing set cover on the set of residues and on the undirected graph that
describes residue exchangeability given the set cover. For fixed functional
forms indicating how to obtain edge weights from the set cover and, after that,
substitution-matrix elements from weighted distances on the graph, the resulting
substitution matrix can be checked for performance against some known set of
reference alignments and for given gap costs. Finding the appropriate functional
forms and gap costs can then be formulated as an optimization problem that seeks
to maximize the performance of the substitution matrix on the reference
alignment set. We give computational results on the BAliBASE suite using a
genetic algorithm for optimization. Our results indicate that it is possible to
obtain substitution matrices whose performance is either comparable to or
surpasses that of several others, depending on the particular scenario under
consideration.

\bigskip
\noindent
{\bf Keywords:} Sequence alignment, Substitution matrix, Residue set cover.
\end{abstract}

\section{Introduction}\label{intro}

One of the most central problems of computational molecular biology is to align
two sequences of residues, a residue being generically understood as a
nucleotide or an amino acid, depending respectively on whether the sequences
under consideration are nucleic acids or proteins. This problem lies at the
heart of several higher-level applications, such as heuristically searching
sequence bases \cite{lp85,pl88,agmml90,amszzml97} or aligning a larger number
of sequences concomitantly \cite{g97,sm97,g99,p00,ltptp01,n02,ejt04} for the
identification of special common substructures (the so-called motifs, cf.\ 
\cite{d81,bjeg98,vms99,rfpgp00,sw03}) that encode structural or functional
similarities of the sequences \cite{t99,vvp01,b02,l03} or yet the sequences'
promoter regions in the case of nucleic acids \cite{vms99}, for example.

Finding the best alignment between two sequences is based on maximizing a
scoring function that quantifies the overall similarity between the sequences.
Normally this similarity function has two main components. The first one is a
symmetric matrix, known as the substitution matrix for the set of residues under
consideration, which gives the contribution the function is to incur when two
residues are aligned to each other. The second component represents the cost of
aligning a residue in a sequence to a gap in the other, and gives the negative
contribution to be incurred by the similarity function when this happens. There
is no consensually accepted, general-purpose criterion for selecting a
substitution matrix or a gap-cost function. Common criteria here include those
that stem from structural or physicochemical characteristics of the residues
(e.g., \cite{f66,g74,mmy79,fjd85,r87,dt01}) and those that somehow seek to
reproduce well-known alignments as faithfully as possible (e.g.,
\cite{m71,dso78,gb86,lrg86,rddh88,gcb92,hh92,jtt92,jo93,bcg94, rssbs97,jl00,
mv00,bc01,lmt01, hr02,msv02,vst03,xm04}). Useful surveys include
\cite{vw94,hh00,v02}.

We then see that, even though an optimal alignment between two sequences is
algorithmically well understood and amenable to being computed efficiently, the
inherent difficulty of selecting appropriate scoring parameters suggests that
the problem is still challenging in a number of ways. This is especially true of
the case of protein alignment, owing primarily to the fact that the set of
residues is significantly larger than in the case of nucleic acids, and also to
the existence of a multitude of criteria whereby amino acids can be structurally
or functionally exchanged by one another.

For a given structural or physicochemical property (or set of properties) of
amino acids, this exchangeability may be expressed by a set cover of the set of
all amino acids, that is, by a collection of subsets of that set that includes
every amino acid in at least one subset. Each of these subsets represents the
possibility of exchanging any of its amino acids by any other. Set covers in
this context have been studied extensively \cite{s66,jz81,t86,ss90,lb93,m95,
k96,nfjwn96,s96,wb96,i97,vb01,ctrv02,lfww03} and constitute our departing point
in this paper. As we describe in Section~\ref{meth}, we introduce a new
methodology for discovering both an appropriate substitution matrix and gap-cost
parameters that starts by considering an amino-acid set cover. It then builds a
graph from the set cover and sets up an optimization problem whose solution is
the desired substitution matrix and gap costs.\footnote{Our new methodology is
ultimately related to the work of several other authors that have dealt with the
issue of assessing the efficacy of a substitution matrix or its relation to
possible groupings of amino acids. We refer the interested reader to
\cite{hh93,vea95,m99,gb02,kgb04}, for example.}

The resulting optimization problem is defined on a set of target sequence pairs,
preferably one that embodies as great a variety of situations as possible. The
target pairs are assumed to have known alignments, so the optimal solution to
the problem of finding parameters comprises the substitution matrix and the gap
costs whose use in a predefined alignment algorithm yields alignments of the
target pairs that in some sense come nearest the known alignments of the same
pairs. Our optimization problem is set up as a problem of combinatorial search,
being therefore highly unstructured and devoid of any facilitating
differentiability properties. Reasonable ways to approach its solution are then
all heuristic in nature. In Section~\ref{res}, we present the results of
extensive computational experiments that employ an evolutionary algorithm and
targets the BAliBASE pairs of amino-acid sequences  \cite{tpp99a,bttp01}.

Notice, in the context of the methodology categorization we mentioned earlier
in passing, that our new methodology is of a dual character: it both relies on
structural and physicochemical similarities among amino acids and depends on a
given set of aligned sequences in order to arrive at a substitution matrix and
gap costs. We return to this hybrid aspect of our methodology in
Section~\ref{concl}, where conclusions are given.

\section{The methodology}\label{meth}

We describe our methodology for sequences on a generic set $R$ of residues and
only specialize it to the case of proteins in Section~\ref{res}. Given two
residue sequences $X$ and $Y$ of lengths $x$ and $y$, respectively, a global
alignment of $X$ and $Y$ can be expressed by the $2\times z$ matrix $A$ having
the property that its first line, when read from left to right, is $X$ possibly
augmented by interspersed gaps, the same holding for the second line and $Y$, so
long as no column of $A$ comprises gaps only. It follows that $z\ge x,y$. In the
case of a local alignment, that is, an alignment of a subsequence of $X$ and
another of $Y$, this matrix representation remains essentially unchanged,
provided of course that $x$ and $y$ are set to indicate the sizes of the two
subsequences.

For a given substitution matrix $S$ and a pair $(h,g)$ of gap
costs,\footnote{For $k>0$, we assume the customary affine function $p(k)=h+gk$
with $h,g>0$ to express the cost of aligning the $k$th gap of a contiguous group
of gaps in a line of $A$ to a residue in the other line as $p(k)-p(k-1)$,
assuming $p(0)=0$ \cite{sm97}.} the similarity score of alignment $A$, denoted
by
$F_S^{h,g}(A)$, is given by
\begin{equation}
\label{score}
F_S^{h,g}(A)=\sum_{j=1}^{z}f_S^{h,g}(A(1,j),A(2,j)),
\end{equation}
where $f_S^{h,g}(A(1,j),A(2,j))$ gives the contribution of aligning $A(1,j)$
to $A(2,j)$ as either $S(A(1,j),A(2,j))$, if neither $A(1,j)$ nor $A(2,j)$ is a
gap; or $-(h+g)$, if either $A(1,j)$ or $A(2,j)$ is the first gap in a
contiguous group of gaps; or yet $-g$, if either $A(1,j)$ or $A(2,j)$ is the
$k$th gap in a contiguous group of gaps for $k>1$. An optimal global alignment
of $X$ and $Y$ is one that maximizes the similarity score of (\ref{score}) over
all possible global alignments of the two sequences. An optimal local alignment
of $X$ and $Y$, in turn, is the optimal global alignment of the subsequences of
$X$ and $Y$ for which the similarity score is maximum over all pairs of
subsequences of the two sequences. The set of all optimal alignments of $X$ and
$Y$ may be exponentially large in $x$ and $y$, but it does nonetheless admit a
concise representation as a matrix or directed graph that can be computed
efficiently by well-known dynamic programming techniques
\cite{w83,bw84,nb94,c98}, regardless of whether a global alignment of the two
sequences is desired \cite{nw70} or a local one \cite{sw81}. We refer to this
representation as $\mathcal{A}_{X,Y}^*$.

Our strategy for the determination of a suitable substitution matrix starts with
a set cover $\mathcal{C}=\{C_1,\ldots,C_c\}$ of the residue set $R$, that is,
$\mathcal{C}$ is such that $C_1\cup\cdots\cup C_c=R$. Next we define $G$ to be
an undirected graph of node set $R$ having an edge between two nodes (residues)
$u$ and $v$ if and only if at least one of $C_1,\ldots,C_c$ contains both $u$
and $v$. Graph $G$ provides a natural association between how exchangeable a
node is by another and the distance between them in the graph. Intuitively, the
closer two nodes are to each other in $G$ the more exchangeable they are and we
expect an alignment of the two to contribute relatively more positively to the
overall similarity score. Quantifying this intuition involves crucial decisions,
so we approach the problem in two careful steps, each leaving considerable room
for flexibility. The first step consists of turning $G$ into a weighted graph,
that is, assigning nonnegative weights to its edges, and then computing the
weighted distance between all pairs of nodes.\footnote{Weight non-negativity is
crucial here, since the presence of negative weights may render the
weighted-distance problem ill-posed \cite{clrs01}.} The second step addresses
the turning of these weighted distances into elements of a substitution matrix
so that larger distances signify ever more restricted exchangeability.

Let us begin with the first step. For $(u,v)$ an edge of $G$, let $w(u,v)$
denote its weight. We define the value of $w(u,v)$ on the premise that, if the
exchangeability of $u$ and $v$ comes from their concomitant membership in a
large set of $\mathcal{C}$, then it should eventually result in a smaller
contribution to the overall similarity score than if they were members of a
smaller set. In other words, the former situation bespeaks an intuitive
``weakness'' of the property that makes the two residues exchangeable. In broad
terms, then, we should let $w(u,v)$ be determined by the smallest of the sets of
$\mathcal{C}$ to which both $u$ and $v$ belong, and should also let it be a
nondecreasing function of the size of this smallest set.

Let $c^-$ be the size of the smallest set of $\mathcal{C}$ and $c^+$ the size of
its largest set. Let $c_{u,v}^-$ be the size of the smallest set of
$\mathcal{C}$ of which both $u$ and $v$ are members. We consider two functional
forms according to which $w(u,v)$ may depend on $c_{u,v}^-$ as a nondecreasing
function. Both forms force $w(u,v)$ to be constrained within the interval
$[w^-,w^+]$ with $w^-\ge 0$. For $\lambda\ge 1$, the first form is the convex
function
\begin{equation}
\label{wf1}
w_1(u,v)=w^-+(w^+-w^-)\left(\frac{c_{u,v}^--c^-}{c^+-c^-}\right)^\lambda,
\end{equation}
while the second is the concave function
\begin{equation}
\label{wf2}
w_2(u,v)=w^+-(w^+-w^-)\left(\frac{c^+-c_{u,v}^-}{c^+-c^-}\right)^\lambda.
\end{equation}
Having established weights for all the edges of $G$, let $d_{u,v}$ denote the
weighted distance between nodes $u$ and $v$. Clearly, $d_{u,u}=0$ and, if no
path exists in $G$ between $u$ and $v$ (i.e., $G$ is not connected and the two
nodes belong to two different connected components), then $d_{u,v}=\infty$.

Carrying out the second step, which is obtaining the elements of the
substitution matrix from the weighted distances on $G$, involves difficult
choices as well. While, intuitively, it is clear that residues separated by
larger weighted distances in $G$ are to be less exchangeable for each other than
residues that are closer to each other (in weighted terms) in $G$, the
functional form that the transformation of weighted distances into
substitution-matrix elements is to take is once again subject to somewhat
arbitrary decisions. What we do is to set $S(u,v)=0$ if $d_{u,v}=\infty$, and to
consider two candidate functional forms for the transformation in the case of
finite distances.

Let us initially set $[S^-,S^+]$ as the interval within which each element of
the substitution matrix $S$ is to be constrained (we assume $S^->0$ for
consistency with the substitution-matrix element that goes with an infinite
distance, whose value we have just set to $0$). Let us also denote by $d^+$ the
largest (finite) weighted distance occurring in $G$ for the choice of weights at
hand. We then consider two functional forms for expressing the dependency of
$S(u,v)$, as a nonincreasing function, upon a finite $d_{u,v}$. For $\mu\ge 1$,
we once again consider a convex function,
\begin{equation}
\label{sf1}
S_1(u,v)=S^-+(S^+-S^-)\left(\frac{d^+-d_{u,v}}{d^+}\right)^\mu,
\end{equation}
and a concave one,
\begin{equation}
\label{sf2}
S_2(u,v)=S^+-(S^+-S^-)\left(\frac{d_{u,v}}{d^+}\right)^\mu.
\end{equation}

\begin{figure}
\centering
\begin{tabular}{c@{\hspace{0.00in}}c}
\scalebox{0.350}{\includegraphics{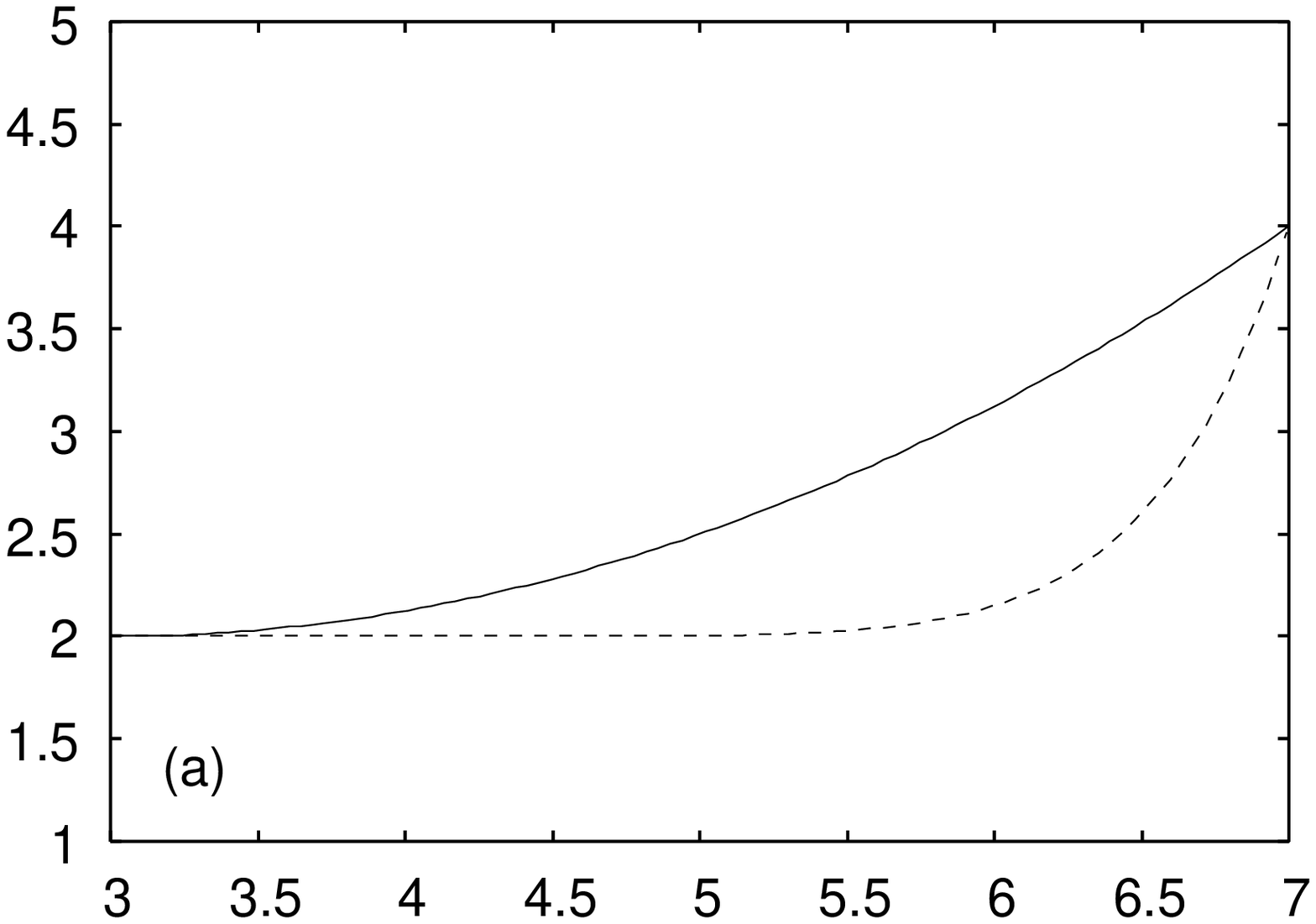}}&
\scalebox{0.350}{\includegraphics{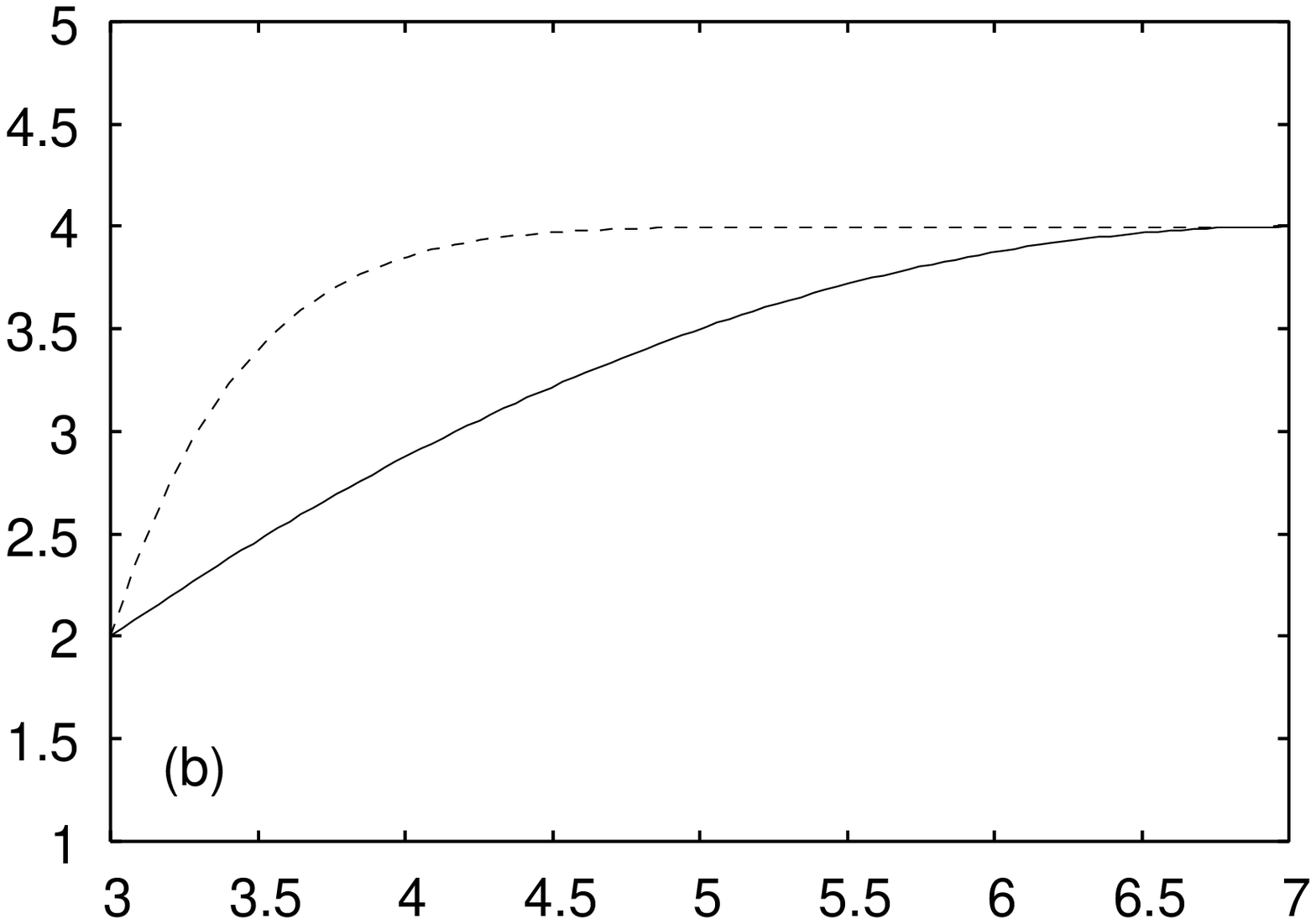}}\\
\scalebox{0.350}{\includegraphics{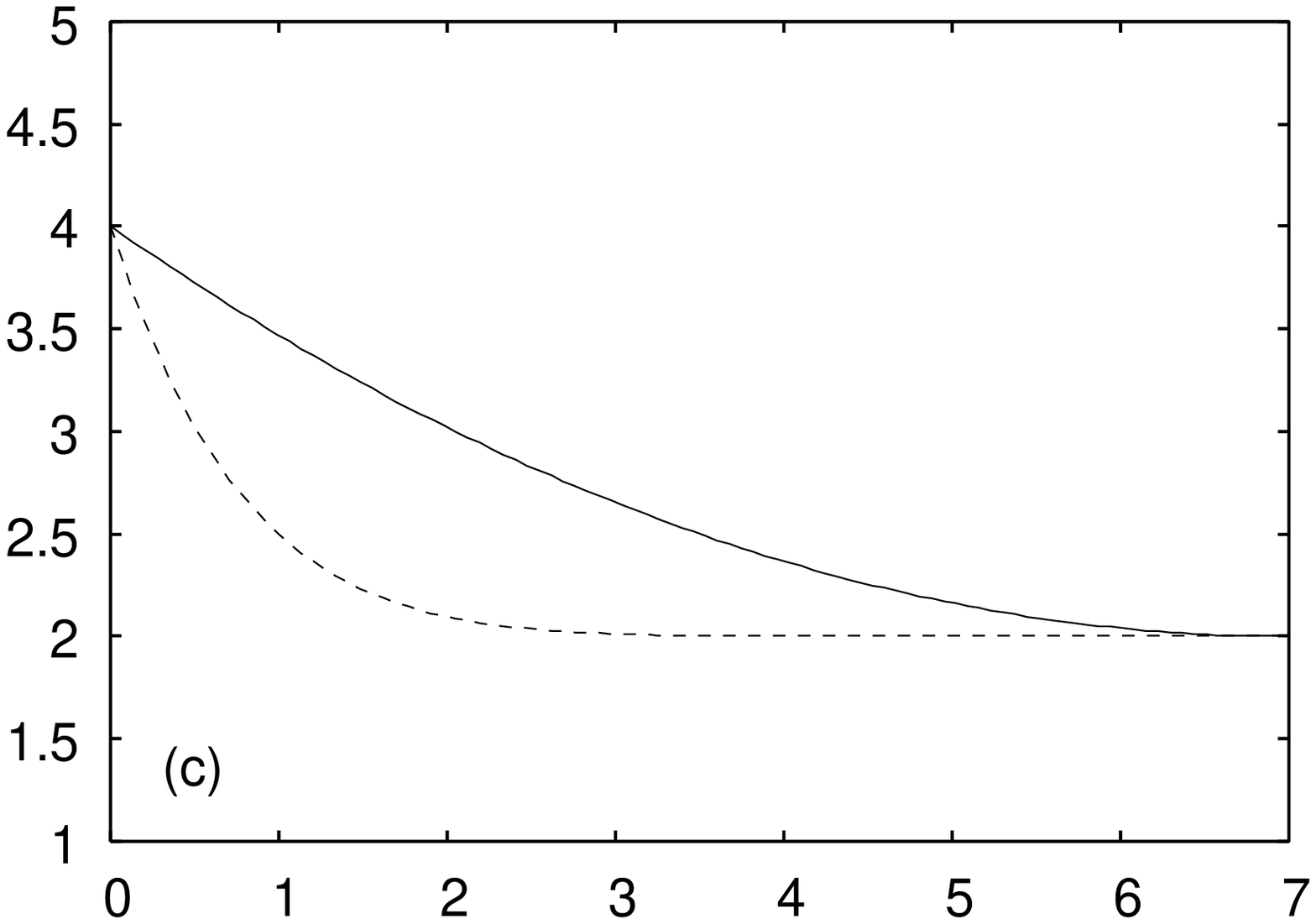}}&
\scalebox{0.350}{\includegraphics{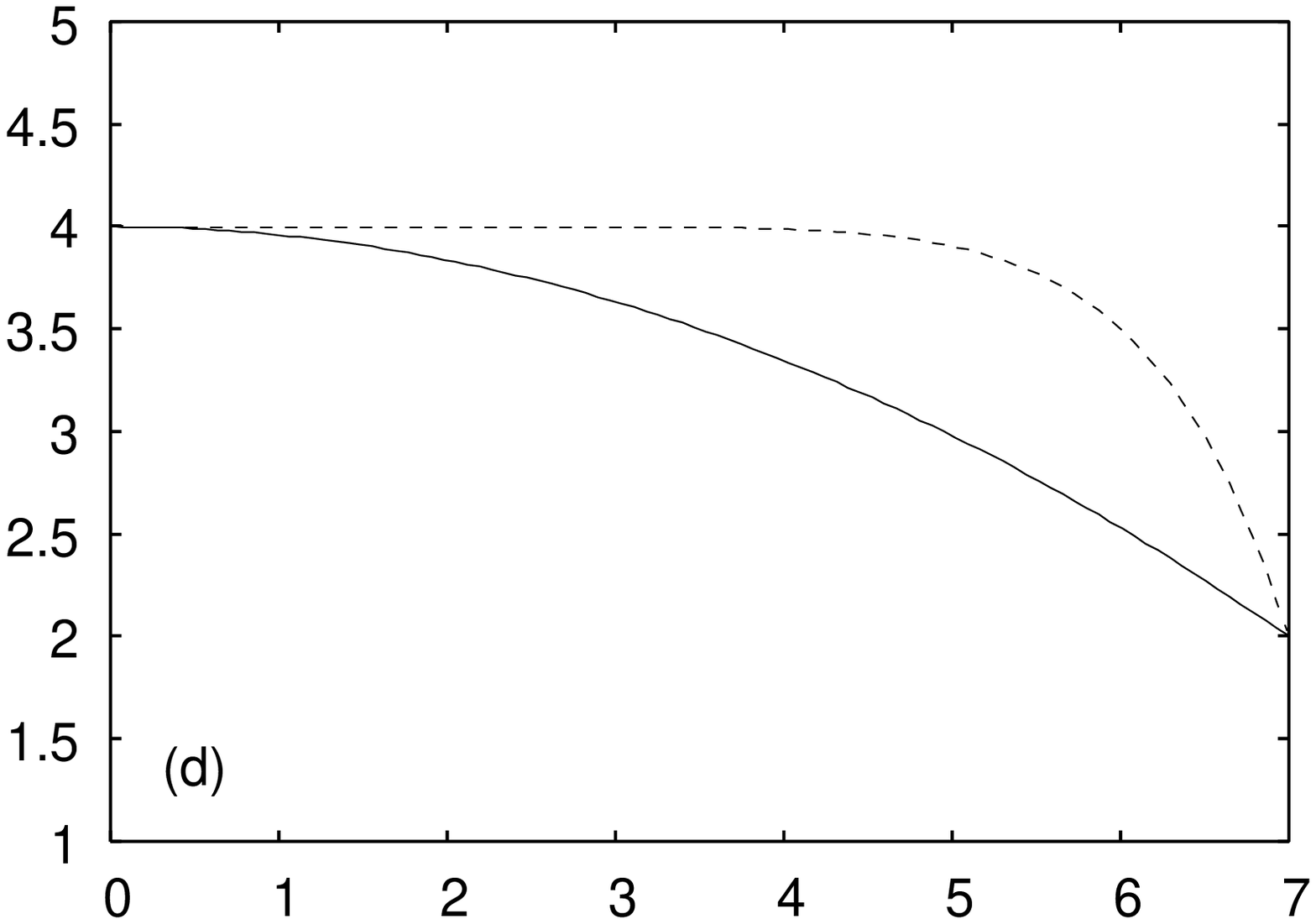}}
\end{tabular}
\caption{Illustrative plots for (\ref{wf1})--(\ref{sf2}), respectively in
(a)--(d). In (a) and (b), $w^-=2$, $w^+=4$, $c^-=3$, $c^+=7$, and $\lambda=2$
(solid plot) or $\lambda=9$ (dashed plot). In (c) and (d), $S^-=2$, $S^+=4$,
$d^+=7$, and $\mu=2$ (solid plot) or $\mu=9$ (dashed plot).}
\label{curves}
\end{figure}

In Figure~\ref{curves} we provide examples of the candidate functional forms for
$w_1(u,v)$, $w_2(u,v)$, $S_1(u,v)$, and $S_2(u,v)$ as given by
(\ref{wf1})--(\ref{sf2}), respectively. Each functional form is illustrated for
two $\lambda$ or $\mu$ values, as the case may be.

Once we decide on one of the two functional forms (\ref{wf1}) or (\ref{wf2}),
and similarly on one of (\ref{sf1}) or (\ref{sf2}), and also choose values for
$w^-$, $w^+$, $\lambda$, $S^-$, $S^+$, and $\mu$, then the substitution matrix
$S$ as obtained from $\mathcal{C}$ is well-defined and, together with the
gap-cost parameters $h$ and $g$, can be used to find the representation
$\mathcal{A}_{X,Y}^*$ of the set of all optimal (global or local) alignments
between the two sequences $X$ and $Y$. The quality of our choices regarding
functional forms and parameters, and hence the quality of the resulting $S$,
$h$, and $g$, can be assessed if a reference alignment, call it $A_{X,Y}^r$, is
available for the two sequences. When this is the case, we let
$\rho_S^{h,g}(A_{X,Y}^r,\mathcal{A}_{X,Y}^*)$ be the fraction of the columns of
$A_{X,Y}^r$ that also appear in at least one of the alignments that are
represented in $\mathcal{A}_{X,Y}^*$.\footnote{This definition must be read with
care. If a certain column of $A_{X,Y}^r$ refers to a certain occurrence of
residue $\alpha$ in $X$ and of residue $\beta$ in $Y$, then it counts towards
$\rho_S^{h,g}(A_{X,Y}^r,\mathcal{A}_{X,Y}^*)$ only if the same two occurrences
of $\alpha$ and $\beta$ are aligned to each other in at least one of the
alignments represented in $\mathcal{A}_{X,Y}^*$. The cases in which a residue in
one of the two sequences is aligned to a gap in the other in $A_{X,Y}^r$ are
entirely analogous. The required bookkeeping in any of the cases is simple to
perform if one resorts to the matrix or directed graph that gives the structure
of $\mathcal{A}_{X,Y}^*$.} The substitution matrix $S$, and also $h$ and $g$,
are then taken to be as good for $A_{X,Y}^r$ as
$\rho_S^{h,g}(A_{X,Y}^r,\mathcal{A}_{X,Y}^*)$ is close to $1$.

Thus, given a residue set cover $\mathcal{C}$ and a set $\mathcal{A}^r$ of
reference alignments (each alignment on a different pair of sequences over the
same residue set $R$), obtaining the best possible substitution matrix $S$ and
gap-cost parameters $h$ and $g$ can be formulated as the following optimization
problem: find functional forms and parameters that maximize some (for now
unspecified) average of $\rho_S^{h,g}(A_{X,Y}^r,\mathcal{A}_{X,Y}^*)$ over all
pairs $(X,Y)$ of sequences such that $A_{X,Y}^r\in\mathcal{A}^r$. In the next
section, we make this definition precise when residues are amino acids and
proceed to the description of computational results.

\section{Computational results}\label{res}

Let $b_w$ be a two-valued variable indicating which of (\ref{wf1}) or
(\ref{wf2}) is to be taken as the functional form for the edge weights, and
similarly let $b_S$ indicate which of (\ref{sf1}) or (\ref{sf2}) is to give the
functional form for the elements of $S$. These new parameters defined, we begin
by establishing bounds on the domains from which each of the other eight
parameters involved in the optimization problem may take values, and also
make those domains discrete inside such bounds by taking equally spaced
delimiters. For the purposes of our study in this section, this results in what
is shown in Table~\ref{params}.

\begin{table}
\centering
\caption{Parameters and their domains.}
\input{tab1}
\label{params}
\end{table}

The parameter domains shown in Table~\ref{params} make up for over $3.7$
trillion possible combinations, yielding about $1.6$ billion different
substitution matrices.\footnote{This is only a rough estimate, since there are
combinations that yield the same substitution matrix. For example, setting
$\lambda=1$ renders $b_w$ needless, the same holding for $\mu$ and $b_S$. In a
similar vein, setting $w^-=w^+=1$ renders both $b_w$ and $\lambda$ needless, and
similarly for $b_S$ and $\mu$ (together with $b_w$, $w^-$, $w^+$, and $\lambda$)
when $S^-=S^+=1$.} The set of all such combinations seems to be structured in
no usable way, so finding the best combination with respect to some set of
reference alignments as discussed in Section~\ref{meth} must not depend on any
technique of explicit enumeration but rather on some heuristic approach.

The approach we use in this section is to employ an evolutionary algorithm for
finding the best possible combination within reasonable time bounds. Each
individual for this algorithm is a $10$-tuple indicating one of the possible
combination of parameter values. Our evolutionary algorithm is a standard
generational genetic algorithm \cite{m96}. It produces a sequence of
$100$-individual generations, the first of which is obtained by randomly
choosing a value for each of the $10$ parameters in order to produce each of its
individuals. Each of the subsequent generations is obtained from the current
generation by a combination of crossover and mutation operations, following an
initial elitist step whereby the $5$ fittest individuals of the current
generation are copied to the new one. While the new generation is not full,
either a pair of individuals is selected from the current generation to undergo
crossover (with probability $0.5$) or one individual is selected to undergo
a single-locus mutation (with probability $0.5$).\footnote{Both the crossover
point and the locus for mutation are chosen at random, essentially with the
parameters' domains in mind, so that the probability that such a choice singles
out a parameter whose domain has size $a$ is proportional to $\log a$. Mutating
the parameter's value is achieved straightforwardly, while breaking the
$10$-tuples for crossover requires the further step of interpreting the
parameter as a binary number.} The pair of individuals resulting from the
crossover, or the single mutated individual, is added to the new generation,
unless an individual that is being added is identical to an individual that
already exists in the population. When this happens, the duplicating individual
is substituted for by a randomly generated individual. Selection is performed in
proportion to the individuals' linearly normalized fitnesses.\footnote{This
means that, for $1\le k\le 100$, the $k$th fittest individual in the generation
is selected with probability proportional to $L-(L-1)(k-1)/99$, where $L$ is
chosen so that the expression yields a value $L$ times larger for the fittest
individual than it does for the least fit (for which it yields value $1$). We
use $L=10$ throughout.}

The crux of this genetic algorithm is of course how to assess an individual's
fitness, and this is where an extant set of reference alignments $\mathcal{A}^r$
comes in. In our study we take $\mathcal{A}^r$ to be the set of alignments
present in the BAliBASE suite \cite{bttp01}. It contains $167$ families of
amino-acid sequences arranged into eight reference sets. For each family of the
first five reference sets two pieces of reference information are provided: a
multiple alignment of all the sequences in the family and a demarcation of the
relevant motifs given the multiple alignment. Families in the remaining three
reference sets are not provided with motif demarcations, so we refrain from
using them in our experiments, since the fitness function that we use relies on
reference motifs as well. Note that, even though the BAliBASE suite is targeted
at multiple sequence alignments (cf.\ \cite{tpp99b,ls02} for example
applications), each such alignment trivially implies a pairwise alignment for
all sequence pairs in each family and also motif fragments for each pair. Our
set $\mathcal{A}^r$ then comprises every sequence pair from the BAliBASE suite
for which a reference alignment exists with accompanying motif demarcation.

The organization of the BAliBASE suite suggests a host of possibilities for
evaluating the efficacy of a substitution matrix $S$ and of gap-cost parameters
$h$ and $g$. For a pair of sequences $(X,Y)$, whose reference alignment is
$A_{X,Y}^r\in\mathcal{A}^r$, and recalling that $\mathcal{A}_{X,Y}^*$ represents
the set of all optimal alignments of $X$ and $Y$ given $S$, $h$, and $g$, we use
four variants of the $\rho_S^{h,g}(A_{X,Y}^r,\mathcal{A}_{X,Y}^*)$ of
Section~\ref{meth} as the bases of the fitness function to be used by the
genetic algorithm. These are denoted by
$\rho_{S,1}^{h,g}(A_{X,Y}^r,\mathcal{A}_{X,Y}^*)$ through
$\rho_{S,4}^{h,g}(A_{X,Y}^r,\mathcal{A}_{X,Y}^*)$ and differ among themselves as
to which of the columns of the reference alignment are checked to be present in
at least one of the optimal alignments. We let them be as follows:
\begin{itemize}
\item $\rho_{S,1}^{h,g}(A_{X,Y}^r,\mathcal{A}_{X,Y}^*)$ is based on all the
columns of $A_{X,Y}^r$;
\item $\rho_{S,2}^{h,g}(A_{X,Y}^r,\mathcal{A}_{X,Y}^*)$ is based on all the
columns of $A_{X,Y}^r$ that contain no gaps;
\item $\rho_{S,3}^{h,g}(A_{X,Y}^r,\mathcal{A}_{X,Y}^*)$ is based on all the
columns of $A_{X,Y}^r$ that lie within motifs;
\item $\rho_{S,4}^{h,g}(A_{X,Y}^r,\mathcal{A}_{X,Y}^*)$ is based on all the
columns of $A_{X,Y}^r$ that lie within motifs and contain no gaps.
\end{itemize}

These defined, we first average each one of them over $\mathcal{A}^r$ before
combining them into a fitness function. The average that we take is computed in
the indirectly weighted style of \cite{vea95}, which aims at preventing any
family with overly many pairs, or any pair on which $S$, $h$, and $g$ are
particularly effective, from influencing the average too strongly. The weighting
takes place on an array having $10$ lines, one for each of the nonoverlapping
$0.1$-wide intervals within $[0,1]$, and one column for each of the BAliBASE
families. Initially each pair $(X,Y)$ having a reference alignment $A_{X,Y}^r$
in $\mathcal{A}^r$ is associated with the array cell whose column corresponds to
its family and whose line is given by the interval within which the identity
score of the reference alignment $A_{X,Y}^r$ falls. This score is the ratio of
the number of columns of $A_{X,Y}^r$ whose two amino acids are identical to the
number of columns that have no gaps (when averaging
$\rho_{S,3}^{h,g}(A_{X,Y}^r,\mathcal{A}_{X,Y}^*)$ or
$\rho_{S,4}^{h,g}(A_{X,Y}^r,\mathcal{A}_{X,Y}^*)$, only columns that lie within
motifs are taken into account).

For $1\le k\le 4$, we then let $\rho_{S,k}^{h,g}(\mathcal{A}^r)$ be the
following average of $\rho_{S,k}^{h,g}(A_{X,Y}^r,\mathcal{A}_{X,Y}^*)$ over
$\mathcal{A}^r$. First take the average of
$\rho_{S,k}^{h,g}(A_{X,Y}^r,\mathcal{A}_{X,Y}^*)$ for each array cell over the
sequence pairs $(X,Y)$ that are associated with it (cells with no pairs are
ignored). Then $\rho_{S,k}^{h,g}(\mathcal{A}^r)$ is computed by first averaging
those averages that correspond to the same line of the array and finally
averaging the resulting numbers (note that lines whose cells were all ignored
for having no sequence pairs associated with them do not participate in this
final average).

We are then in position to state the definition of our fitness function. We
denote it by $\varphi_S^{h,g}(\mathcal{A}^r)$ to emphasize its dependency on
how well $S$, $h$, and $g$ lead to alignments that are in good accord with the
alignments of $\mathcal{A}^r$. It is given by the standard Euclidean norm of the
four-dimensional vector whose $k$th component is
$\rho_{S,k}^{h,g}(\mathcal{A}^r)$, that is,
\begin{equation}
\label{fitness}
\varphi_S^{h,g}(\mathcal{A}^r)=
\sqrt{
\left[\rho_{S,1}^{h,g}(\mathcal{A}^r)\right]^2+
\cdots+
\left[\rho_{S,4}^{h,g}(\mathcal{A}^r)\right]^2
}.
\end{equation}
Clearly, $0\le \varphi_S^{h,g}(\mathcal{A}^r)\le 2$ always.

The substitution matrices we have used for comparison are shown in
Table~\ref{matrices},\footnote{The denomination \texttt{NWSGAPPEP} is taken from
\cite{gcg91}, whose GCG software package was originally described by
\cite{dhs84}. For global alignments, we use \texttt{BLOSUM62} to refer to a
version of the matrix that has nonnegative elements exclusively (this version is
obtained by adding the absolute value of the least element of the original
matrix to all other elements, provided at least one negative element exists).
The same holds for local alignments, in this case for the matrices
\texttt{BENNER74}, \texttt{GONNET}, and \texttt{PAM250} as well.} where for each
one we give its most common epithet, the reference to where it was originally
described, and, when different from the former, the reference to where the
gap-cost parameters $h$ and $g$ we use with it are to be found for both global
and local alignments. This table is supplemented by Table~\ref{matrixresults},
where for each matrix we show the value of $\varphi_S^{h,g}(\mathcal{A}^r)$ for
both the global- and the local-alignment case; numbers in bold typeface are the
minimum and maximum of the corresponding column. Table~\ref{setcovers} gives the
two set covers we have used: $\mathcal{I}$ is the set cover from \cite{i97},
$\mathcal{S}$ the one from \cite{ss90}.

\begin{table}
\centering
\caption{Substitution matrices used for comparison.}
\input{tab2}
\label{matrices}
\end{table}

\begin{table}
\centering
\caption{Values of $\varphi_S^{h,g}(\mathcal{A}^r)$ for the matrices of
Table~\ref{matrices} under global or local alignments.}
\input{tab3}
\label{matrixresults}
\end{table}

\begin{table}
\centering
\caption{Set covers.}
\input{tab4}
\label{setcovers}
\end{table}

\begin{figure}
\centering
\begin{tabular}{c@{\hspace{0.00in}}c}
\scalebox{0.350}{\includegraphics{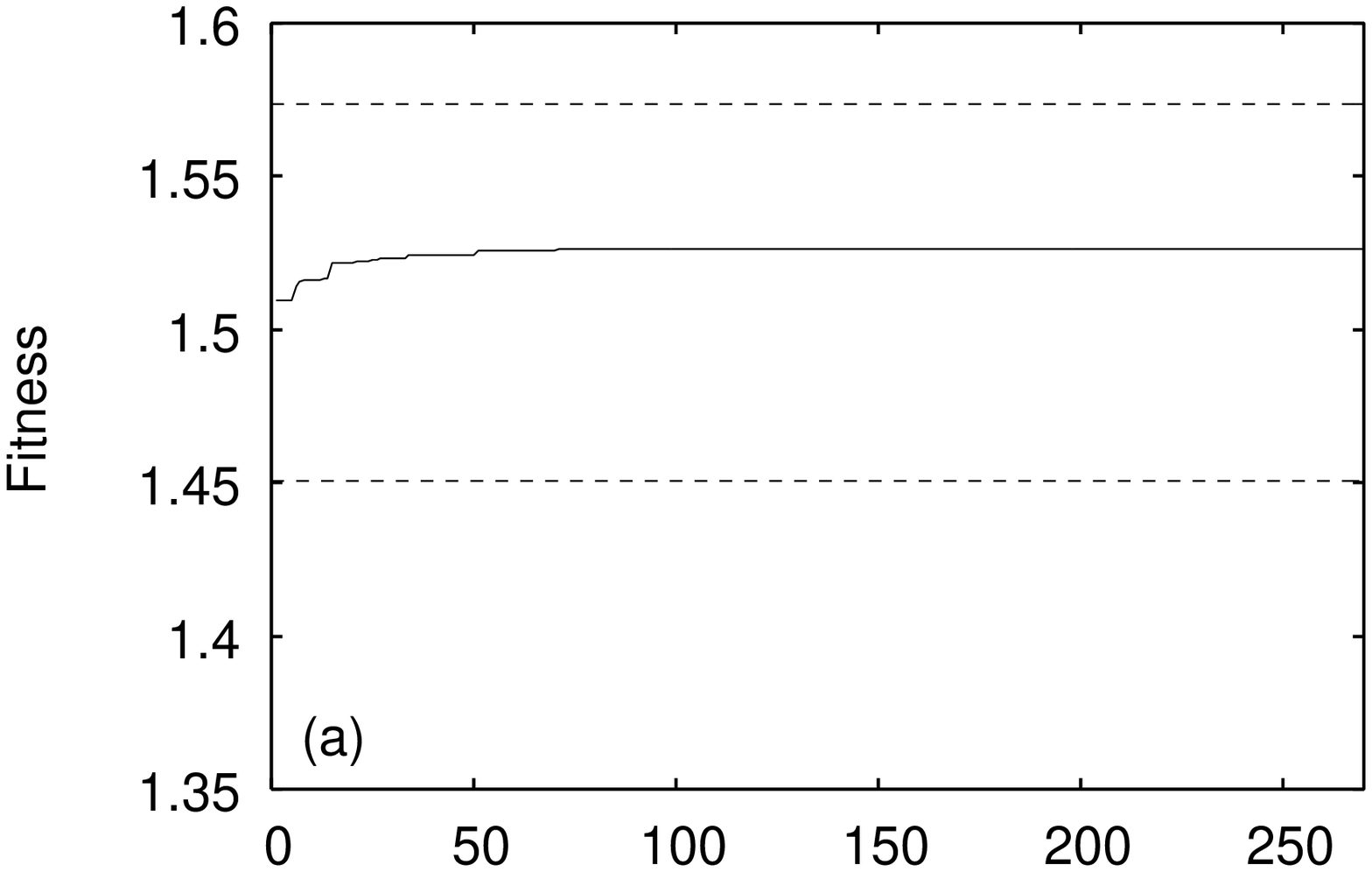}}&
\scalebox{0.350}{\includegraphics{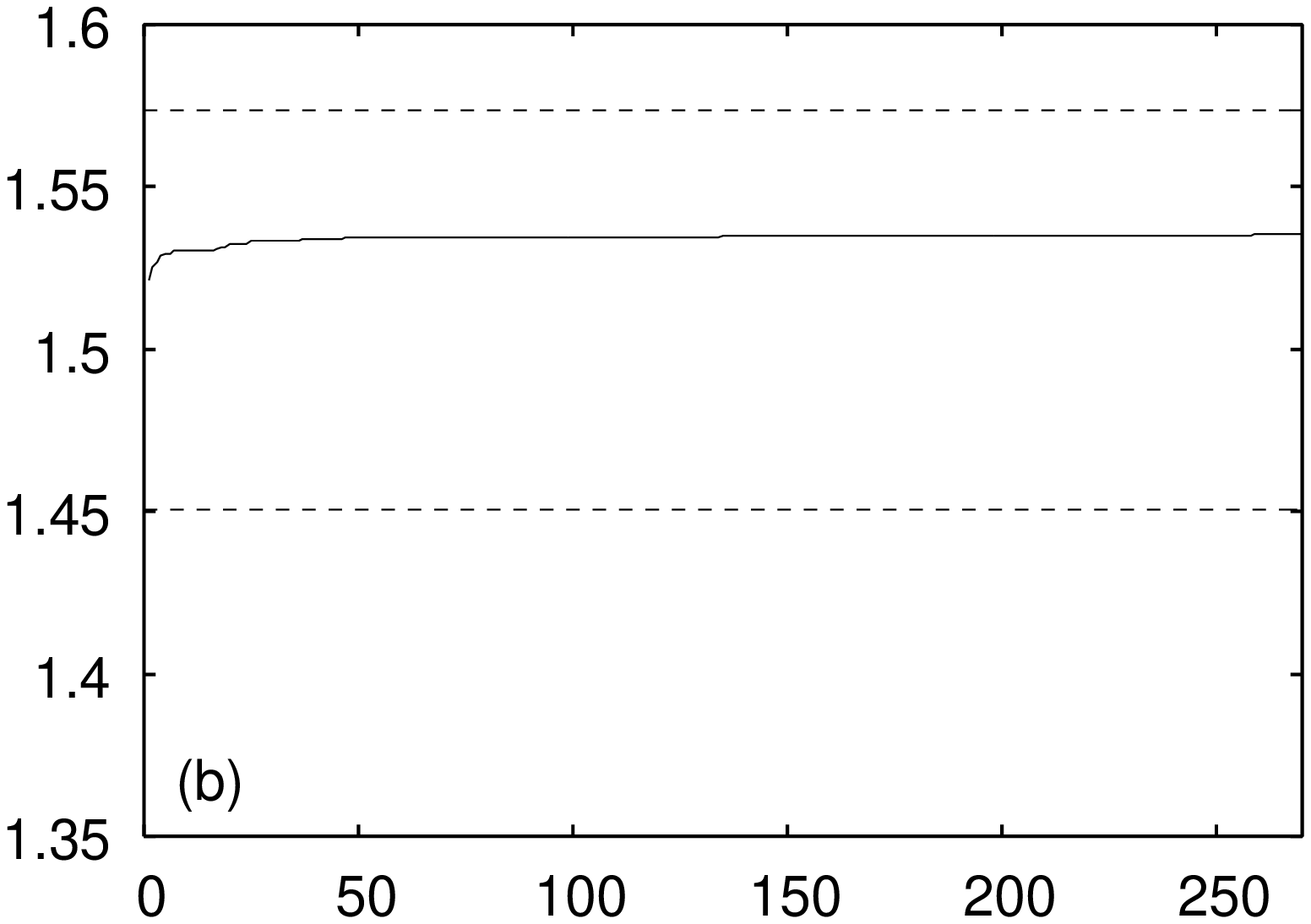}}\\
\scalebox{0.350}{\includegraphics{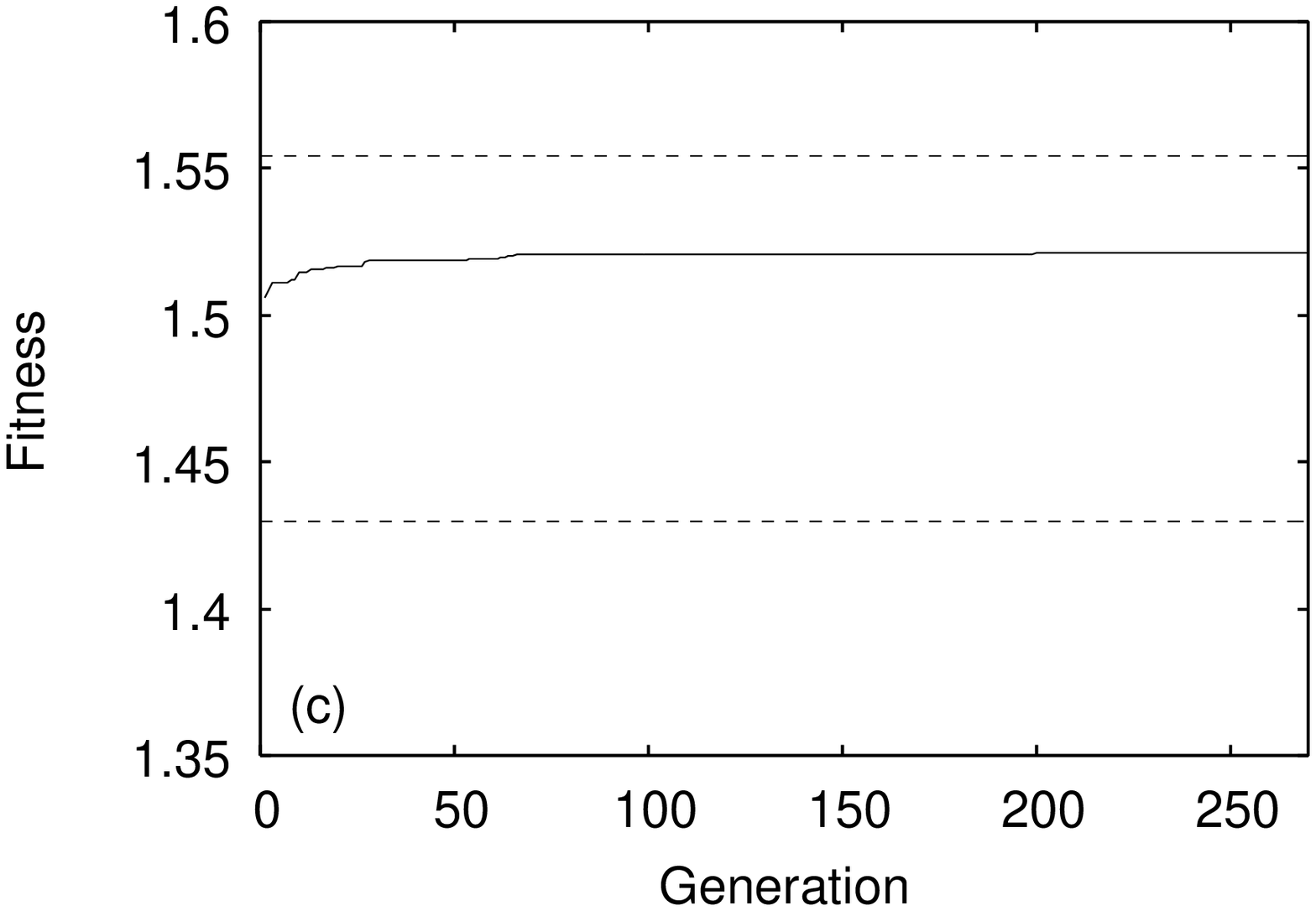}}&
\scalebox{0.350}{\includegraphics{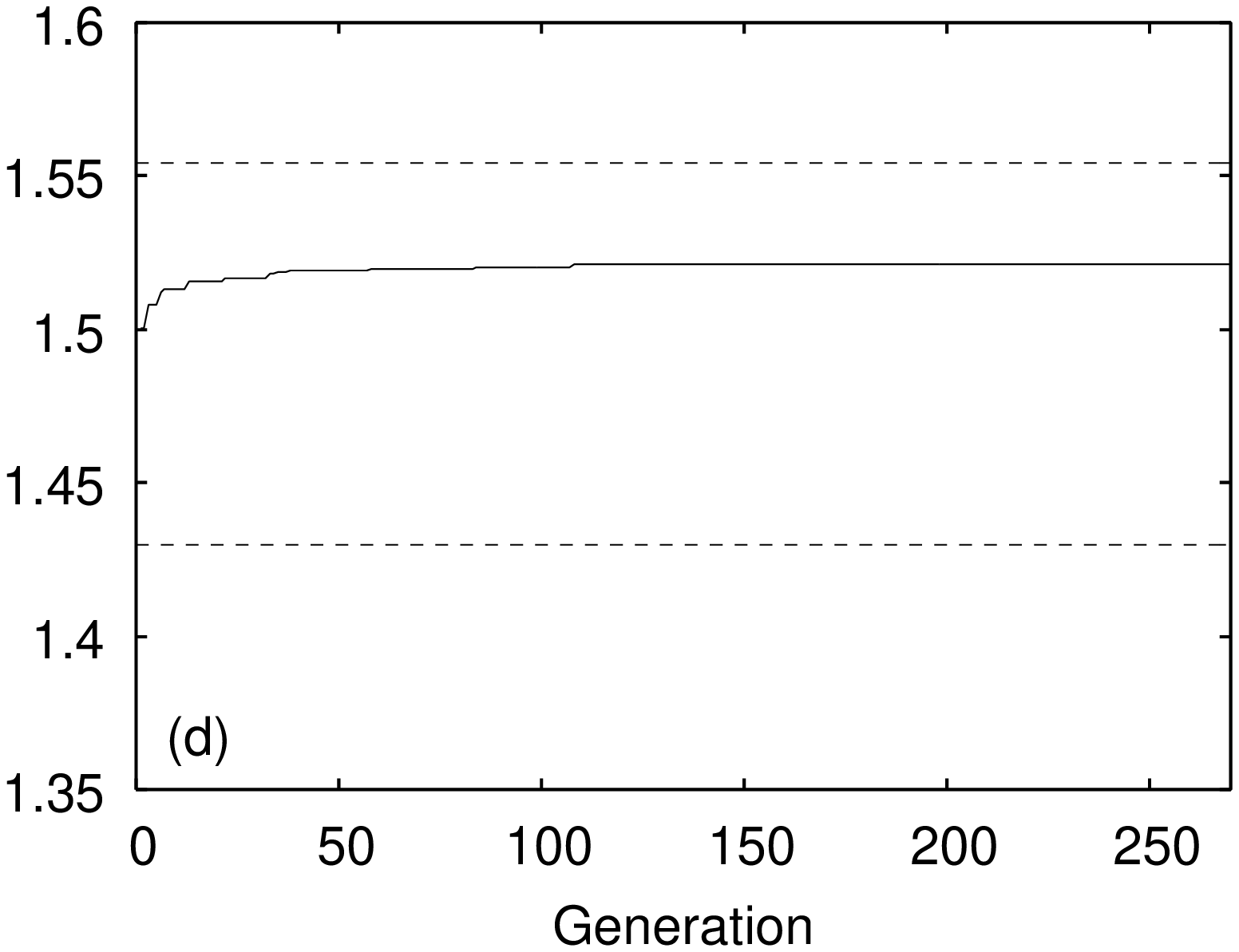}}
\end{tabular}
\caption{Evolution of the fitness, as given by (\ref{fitness}), under global (a
and b) or local (c and d) alignments for the $\mathcal{I}$ (a and c) or
$\mathcal{S}$ (b and d) set covers.}
\label{fitnessevolution}
\end{figure}

One first set of results is summarized in the plots of
Figure~\ref{fitnessevolution} and also in Table~\ref{finalvalues}. Each of the
plots in the figure indicates the evolution of $\varphi_S^{h,g}(\mathcal{A}^r)$
as the genetic algorithm is run for each of the four combinations of global or
local alignments with the $\mathcal{I}$ or $\mathcal{S}$ set cover. At each
generation, what is plotted is the greatest value of
$\varphi_S^{h,g}(\mathcal{A}^r)$ for individuals of that generation, $S$ being
the substitution matrix that corresponds to each individual as explained in
Section~\ref{meth}. We present each plot against two constant values (indicated
as dashed lines) giving the corresponding minimum and maximum highlighted in
Table~\ref{matrixresults}. The best individual of the last generation of each
run is shown as a column in Table~\ref{finalvalues} containing the corresponding
parameter values. Each of Table~\ref{finalvalues}'s columns therefore
corresponds to a substitution matrix, the one output by the corresponding run of
the genetic algorithm, with accompanying gap costs.

\begin{table}
\centering
\caption{Values of the parameters of Table~\ref{params} at the end of each of
the four runs depicted in Figure~\ref{fitnessevolution}. Indications in
parentheses refer to which of parts (a)--(d) of the figure the columns
correspond.}
\input{tab5}

\label{finalvalues}
\end{table}

The first notable feature of the four plots in Figure~\ref{fitnessevolution} is
that, in all cases, the fittest individual of the initial generation is already
well placed with respect to the substitution matrices of Table~\ref{matrices},
even though this generation is the result of a random selection of parameter
values for each of its individuals. This, alone, is in our opinion solid
indication that the essential underlying premise of our new methodology---that
the elements of a substitution matrix can be computed as a function of weighted
distances on the undirected graph that represents a certain amino-acid set
cover---is sound. From the initial generations onward, in all four cases some
rapid progress is made initially, and then fitness improvements become more and
more sporadic. This is no surprise if we consider that the fitness landscape we
are dealing with is completely non-differentiable and probably highly rugged
(i.e., with many local maxima) as well, which is in fact the reason why we give
mutations the high prominence of a $50\%$ chance as a new generation is being
filled.

The question, of course, is whether running the genetic algorithm beyond the
$270$ generations of the figure can lead it to eventually find individuals whose
fitnesses go beyond the uppermost dashed lines in the plots (that is,
individuals that surpass the best-performing matrices on the reference
alignments in $\mathcal{A}^r$). Seemingly, this would require some sort of
phase-transition behavior following the slow progress that the plots depict past
the first $50$ generations or so. While such a behavior is known to occur
relatively often when handling hard, unstructured optimization problems
(cf., e.g., \cite{bc04} for a recent example from combinatorial optimization),
in our case carrying over with the algorithm for each single generation has
required roughly $13$ to $14$ hours,\footnote{These data refer to an Intel
Pentium 4 processor running at $2.26$ GHz.} so at first seeking significant
further improvement does seem impractical.

Notice, however, that practically all of this time consumption is related to
computing $\varphi_S^{h,g}(\mathcal{A}^r)$ for each individual in the current
population. Because this is done in a manner that is fully independent from any
other individual, we can speed the overall computation up nearly optimally by
simply bringing more processors into the effort.\footnote{A finer-grained
opportunity for fully independent parallelism can also be identified if we
recognize that computing $\varphi_S^{h,g}(\mathcal{A}^r)$ in essence boils down
to computing each of $\rho_{S,1}^{h,g}(A_{X,Y}^r,\mathcal{A}_{X,Y}^*)$ through
$\rho_{S,4}^{h,g}(A_{X,Y}^r,\mathcal{A}_{X,Y}^*)$, independently, for every pair
$(X,Y)$ having a reference alignment in $\mathcal{A}^r$. Harnessing this form of
parallelism is infeasible, though, given the current technological reality.}

Our second set of results carries the genetic algorithm well beyond the $270$
generations of Figure~\ref{fitnessevolution}. To this end we employed the
parallel strategy outlined above on four processors, and also concentrated
solely on evolving individuals under global alignments for the $\mathcal{S}$ set
cover. We did, in addition, consider only a subset of $\mathcal{A}^r$, denoted
by $\mathcal{A}^{r,1}$, comprising sequence pairs that are relative to the
BAliBASE reference set~1. In this case, the fitness function to be maximized is
$\varphi_S^{h,g}(\mathcal{A}^{r,1})$, defined as in (\ref{fitness}) when
$\mathcal{A}^{r,1}$ substitutes for $\mathcal{A}^r$. Given these
simplifications, computing through each generation has taken roughly $20$
minutes.

The values of $\varphi_S^{h,g}(\mathcal{A}^{r,1})$ for the substitution matrices
of Table~\ref{matrices} are given in Table~\ref{matrixresults-ref1} for global
alignments only. Notice that this table also contains values for the individual
fitness components
$\rho_{S,1}^{h,g}(\mathcal{A}^{r,1}),\ldots,\rho_{S,4}^{h,g}(\mathcal{A}^{r,1})$
for each matrix; these will be used shortly. In Table~\ref{matrixresults-ref1},
as in Table~\ref{matrixresults}, a bold typeface is used to indicate extremal
values within each of the five numeric columns.

\begin{table}
\centering
\caption{Values of $\varphi_S^{h,g}(\mathcal{A}^{r,1})$ and of
$\rho_{S,1}^{h,g}(\mathcal{A}^{r,1}),\ldots,\rho_{S,4}^{h,g}(\mathcal{A}^{r,1})$
for the matrices of Table~\ref{matrices} under global alignments.}
\input{tab6}
\label{matrixresults-ref1}
\end{table}

\begin{figure}
\centering
\begin{tabular}{c@{\hspace{0.05in}}c}
\scalebox{0.350}{\includegraphics{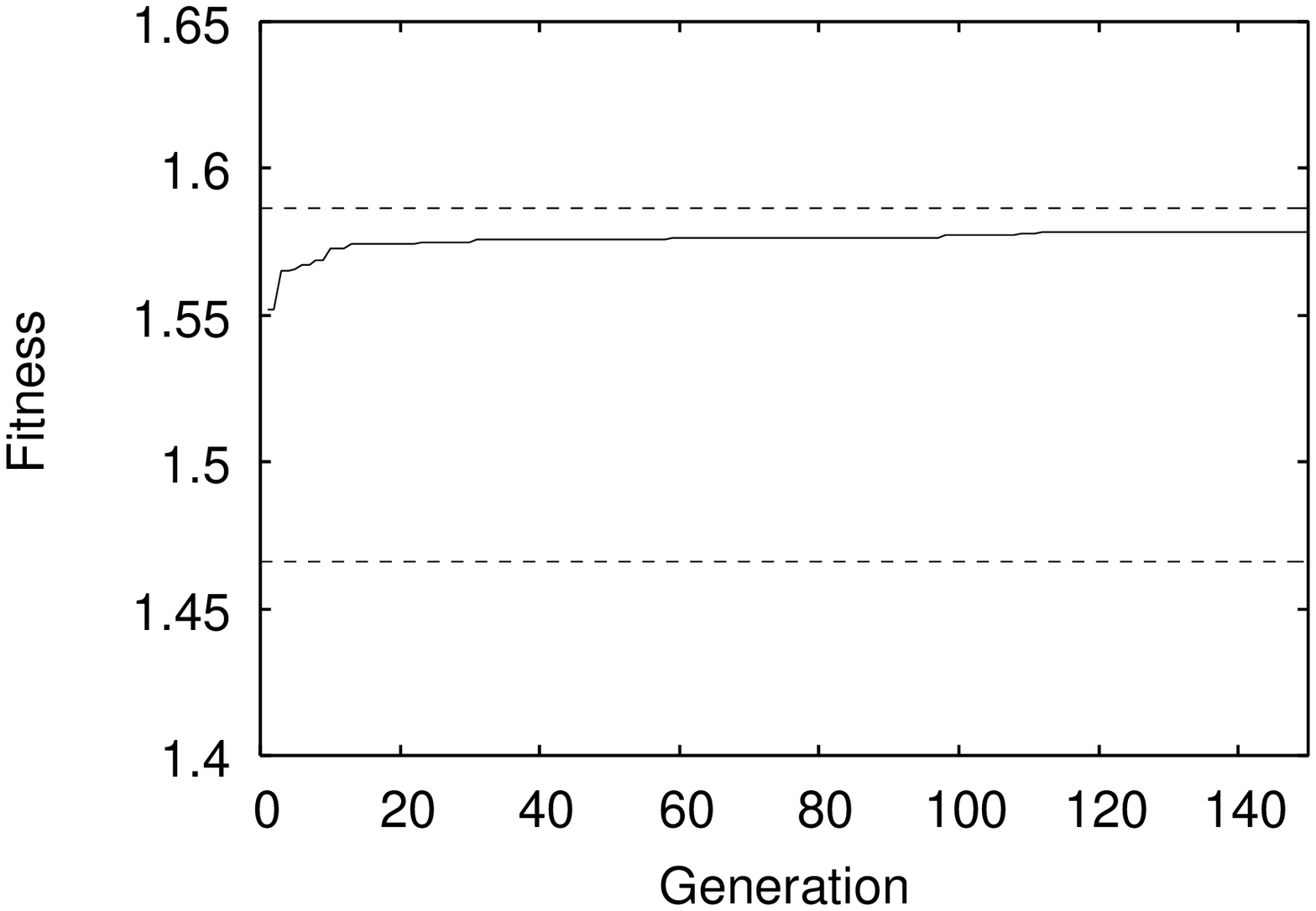}}&
\scalebox{0.350}{\includegraphics{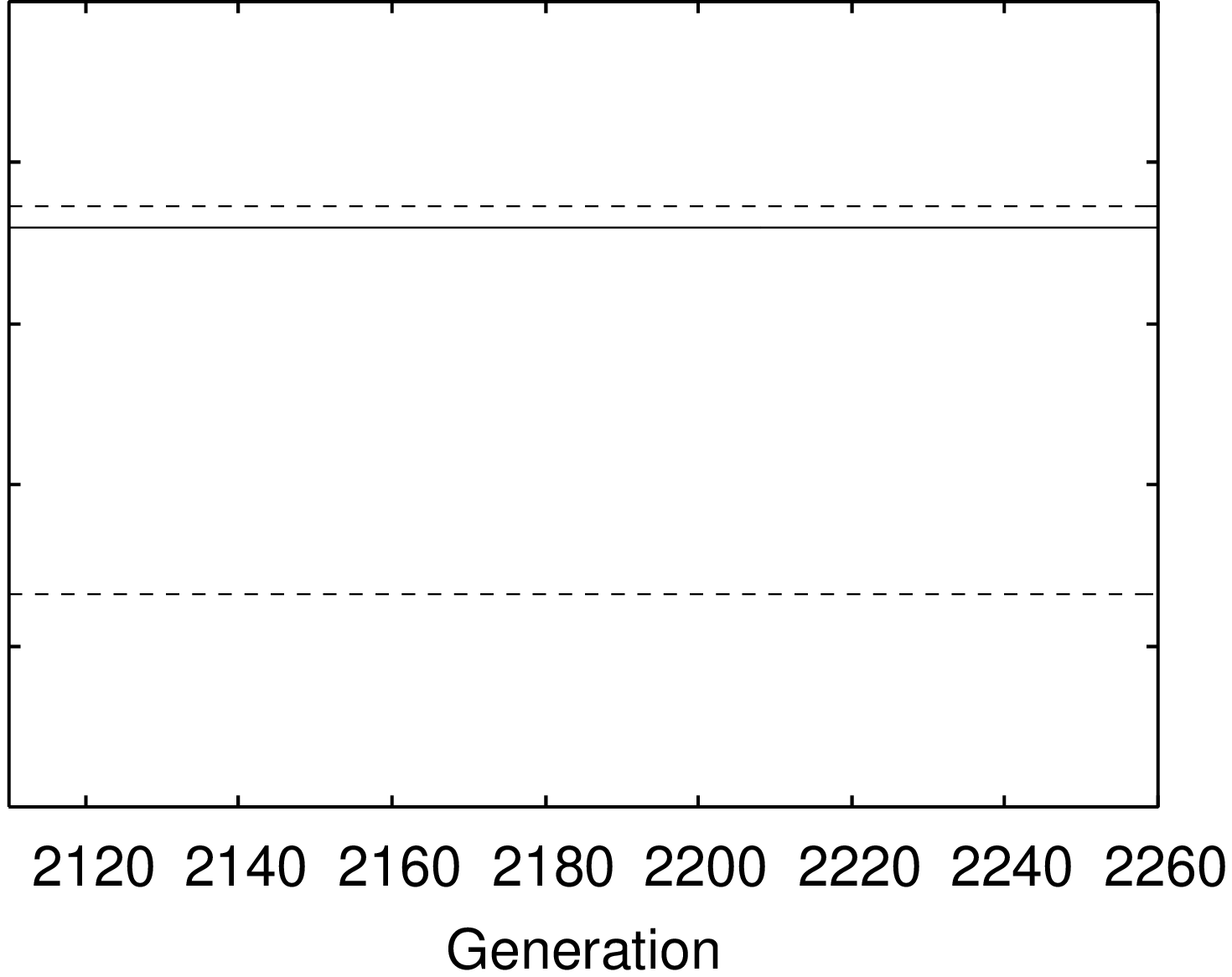}}
\end{tabular}
\caption{Evolution of the fitness, as given by (\ref{fitness}) on
$\mathcal{A}^{r,1}$, under global alignments for the $\mathcal{S}$ set cover.}
\label{fitnessevolution-ref1}
\end{figure}

Figure~\ref{fitnessevolution-ref1} and Table~\ref{finalvalues-ref1} summarize
the results of this smaller-scale experiment. The plot in
Figure~\ref{fitnessevolution-ref1} is analogous to each of the plots in
Figure~\ref{fitnessevolution} and, like them, is given against the dashed lines
that indicate the values highlighted in the leftmost numeric column of
Table~\ref{matrixresults-ref1}. It is presented as two juxtaposed plots on the
initial and final $150$ generations simply for the sake of emphasizing the rapid
fitness growth during the first few tens of generations, on the one hand, and
the very slow growth thereafter, on the other (during the generations that the
plot skips there is growth in one single generation only).
Table~\ref{finalvalues-ref1} is analogous to Table~\ref{finalvalues}, indicating
the parameter values that characterize the fittest individual at the end of the
run of the genetic algorithm.

\begin{table}
\centering
\caption{Values of the parameters of Table~\ref{params} at the end of the run
depicted in Figure~\ref{fitnessevolution-ref1}.}
\input{tab7}

\label{finalvalues-ref1}
\end{table}

\begin{figure}
\centering
\begin{tabular}{c@{\hspace{0.05in}}c}
\scalebox{0.350}{\includegraphics{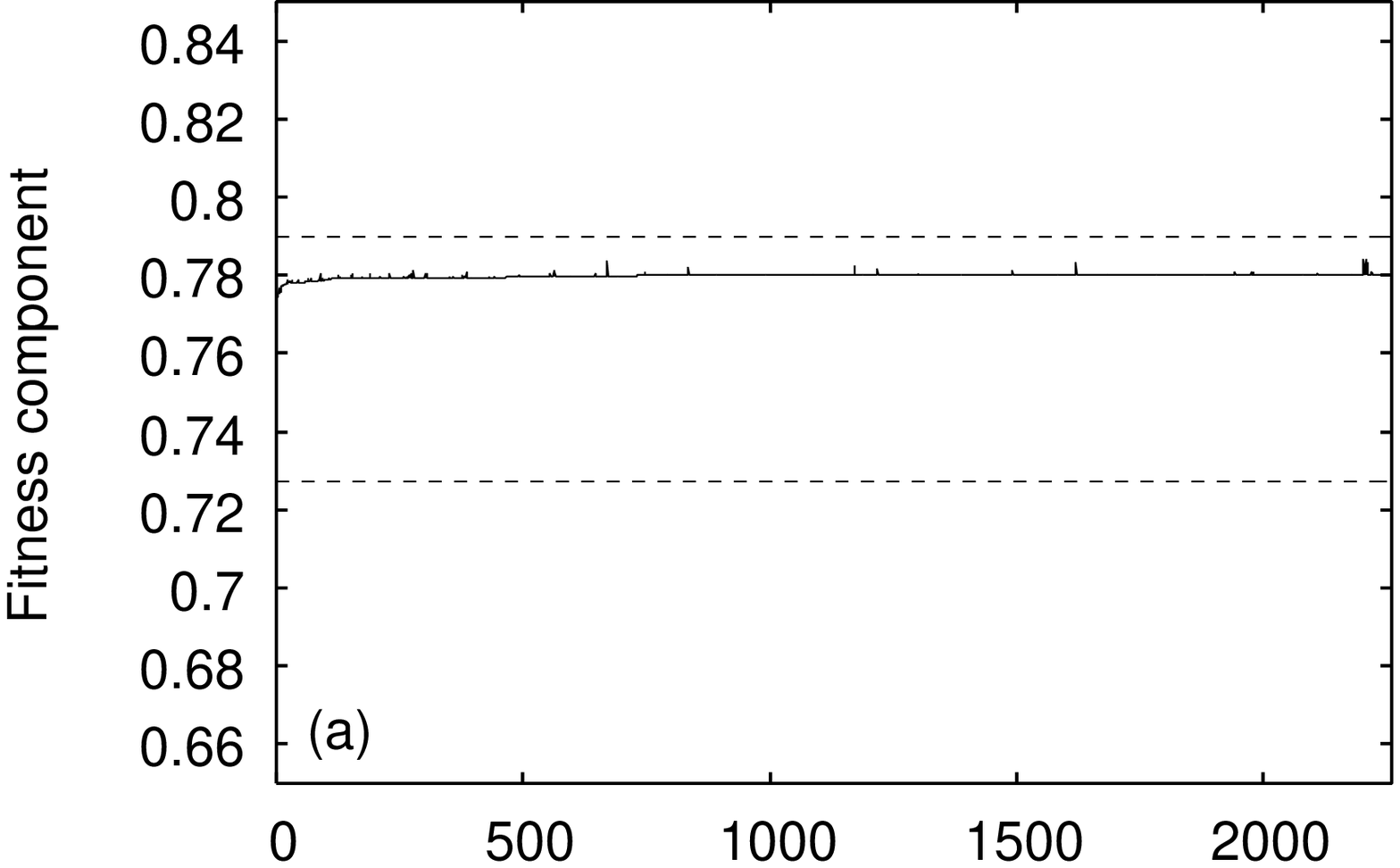}}&
\scalebox{0.350}{\includegraphics{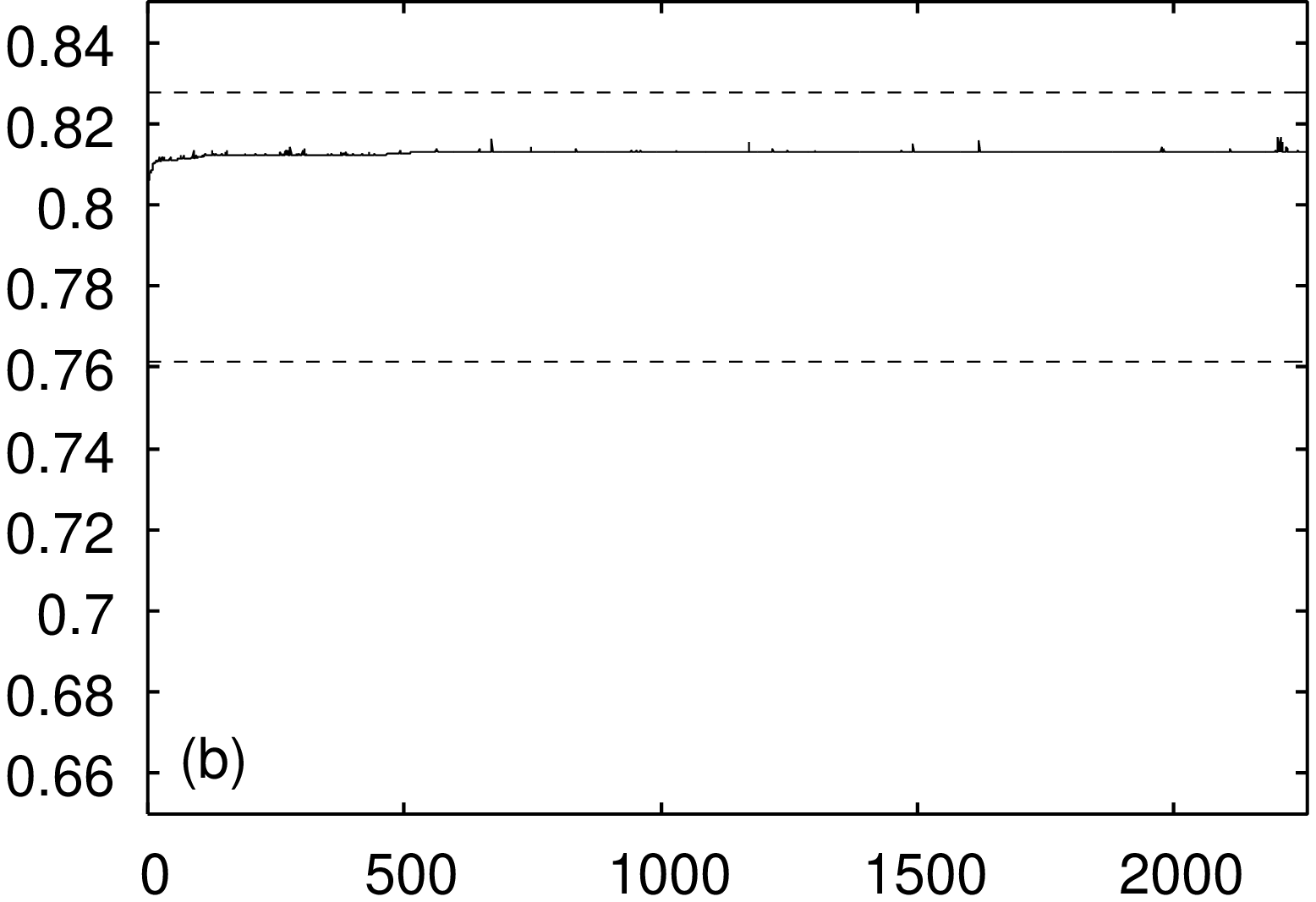}}\\
\scalebox{0.350}{\includegraphics{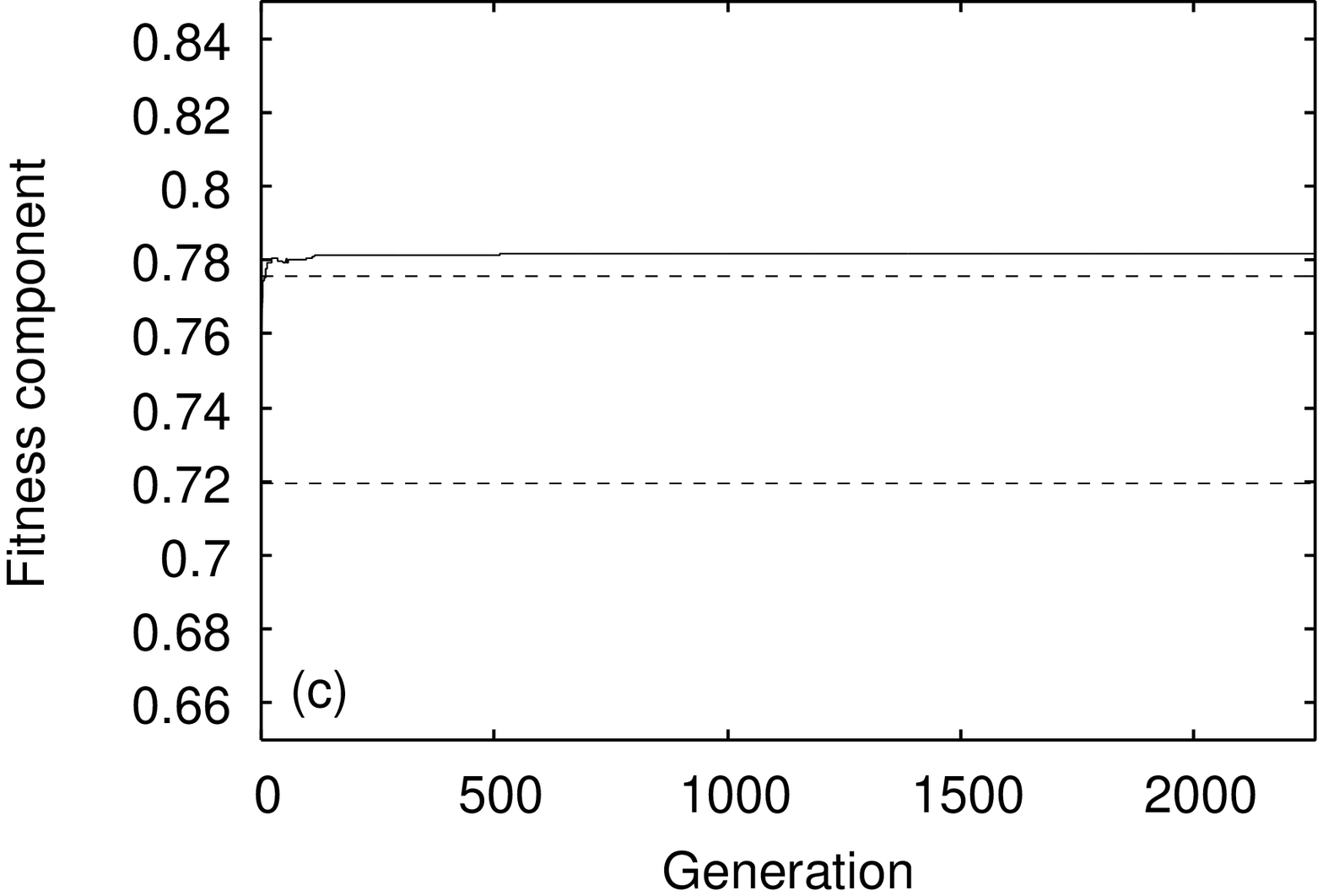}}&
\scalebox{0.350}{\includegraphics{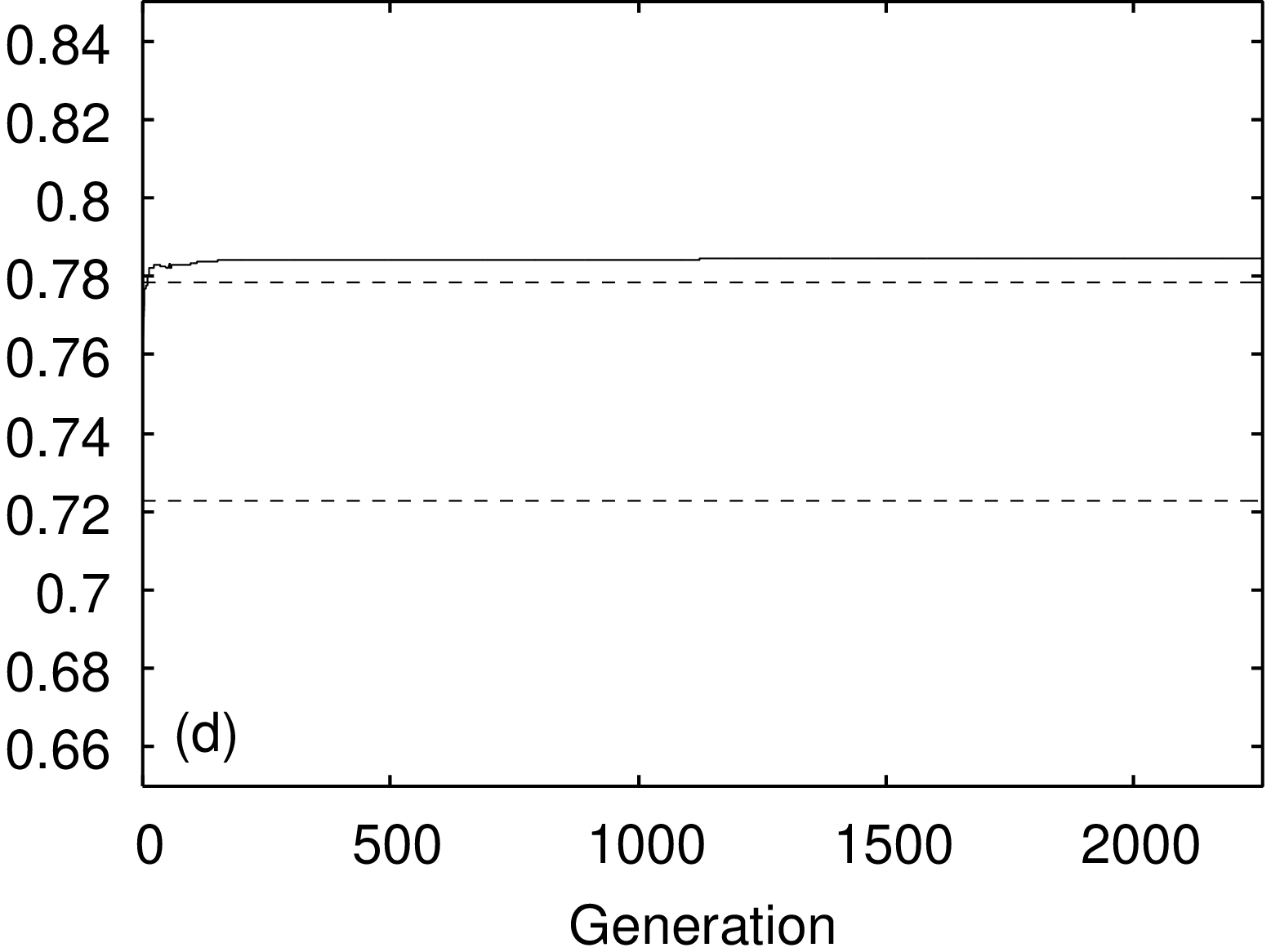}}
\end{tabular}
\caption{Evolution of each of the fitness components
$\rho_{S,1}^{h,g}(\mathcal{A}^{r,1}),\ldots,
\rho_{S,4}^{h,g}(\mathcal{A}^{r,1})$, shown respectively in (a) through (d),
under global alignments for the $\mathcal{S}$ set cover.}
\label{fitnessevolution-ref1-c}
\end{figure}

What is interesting in this second set of results is that, even though nothing
resembling the phase-transition-like behavior alluded to above has taken place,
the fitness of the substitution matrix and gap costs that arise from the
parameter values of Table~\ref{finalvalues-ref1}, specifically $1.5797$, is now
very near $1.5865$, which is the highest value appearing in the leftmost numeric
column of Table~\ref{matrixresults-ref1}. In addition, let us consider the
greatest values of each of
$\rho_{S,1}^{h,g}(\mathcal{A}^{r,1}),\ldots,\rho_{S,4}^{h,g}(\mathcal{A}^{r,1})$
for each generation. Plotting these values against the corresponding minima and
maxima highlighted in the rightmost four columns of
Table~\ref{matrixresults-ref1} yields what is shown in
Figure~\ref{fitnessevolution-ref1-c}, which clearly indicates that the genetic
algorithm very quickly produces a substitution matrix, with associated gap
costs, that surpasses the champion of Table~\ref{matrixresults-ref1} as far as
the fitness components $\rho_{S,3}^{h,g}(\mathcal{A}^{r,1})$ and
$\rho_{S,4}^{h,g}(\mathcal{A}^{r,1})$ are concerned, even though it lags behind
in terms of $\rho_{S,1}^{h,g}(\mathcal{A}^{r,1})$ and
$\rho_{S,2}^{h,g}(\mathcal{A}^{r,1})$. This substitution matrix, it turns out,
is then superior to all the matrices of Table~\ref{matrices} when it comes to
stressing alignment columns that lie within motifs.

\section{Concluding remarks}\label{concl}

We have introduced a new methodology for the determination of amino-acid
substitution matrices. The new methodology starts with a set cover of the
residue alphabet under consideration and builds an undirected graph in which
node vicinity is taken to represent residue exchangeability. The desired
substitution matrix arises as a function of weighted distances in this graph.
Determining the edge weights, and also how to convert the resulting weighted
distances into substitution-matrix elements, constitute the outcome of an
optimization process that runs on a set of reference sequence alignments and
also outputs gap costs for use with the substitution matrix. Our methodology is
then of a hybrid nature: it relies both on the structural and physicochemical
properties that underlie the set cover in use and on an extant set of reference
sequence alignments.

The optimization problem to be solved is well-defined: given parameterized
functional forms for turning cover sets into edge weights and weighted distances
into substitution-matrix elements, the problem asks for parameter values and
gap costs that maximize a certain objective function on the reference set of
alignments. We have reported on computational experiments that use a genetic
algorithm as optimization method and the BAliBASE suite as the source of the
required reference alignments. Our results are supportive of the following main
conclusions. First, that the overall methodology is capable of producing
substitution matrices whose performance falls within the same range of a number
of known matrices' even before any optimization is actually performed (i.e.,
based on the random parameter instantiation that precedes the genetic
algorithm); this alone, we believe, singles out our methodology as a principled
way of determining substitution matrices that concentrates all the effort
related to the structure and physicochemical properties of amino acids on the
discovery of an appropriate set cover. Secondly, that there are scenarios for
which the methodology we introduce can already be claimed to yield a
substitution matrix that surpasses all the others against which it was tested.

We have also found that strengthening this latter conclusion so that it holds in
a wider variety of scenarios depends on how efficiently we can run the genetic
algorithm. Fortunately, it appears that it is all a matter of how many
processors can be amassed for the effort, since the genetic procedure is
inherently amenable to parallel processing and highly scalable, too. There is,
of course, also the issue of investigating alternative functional forms and
parameter ranges to set up the optimization problem, and in fact the issue of
considering other objective functions as well. Together with the search for
faster optimization, these issues make for a very rich array of possibilities
for further study.

\subsection*{Acknowledgments}

The authors acknowledge partial support from CNPq, CAPES, and a FAPERJ BBP
grant.

\bibliography{paper}
\bibliographystyle{plain}

\end{document}

%% file: tab1.tex
\begin{tabular}{lll}
\hline
Parameter&Description&Domain\\
\hline
$b_w$&Selects between (\ref{wf1}) and (\ref{wf2})&$\{1,2\}$\\
$w^-$&Least possible edge weight&$\{0.5,0.55,\ldots,1\}$\\
$w^+$&Greatest possible edge weight&$\{1,1.125,\ldots,5\}$\\
$\lambda$&Exponent for use in (\ref{wf1}) or (\ref{wf2})&$\{1,1.125,\ldots,5\}$\\
$b_S$&Selects between (\ref{sf1}) and (\ref{sf2})&$\{1,2\}$\\
$S^-$&Least possible element of $S$&$\{0.5,0.55,\ldots,1\}$\\
$S^+$&Greatest possible element of $S$&$\{1,1.25,\ldots,25\}$\\
$\mu$&Exponent for use in (\ref{sf1}) or (\ref{sf2})&$\{1,1.125,\ldots,5\}$\\
$h$&Initialization gap cost&$\{2,2.5,\ldots,30\}$\\
$g$&Extension gap cost&$\{0.25,0.375,\ldots,5\}$\\
\hline
\end{tabular}

%% file: tab2.tex
\begin{tabular}{lll}
\hline
Matrix&Original reference&Reference for $h$ and $g$\\
\hline
\texttt{BC0030}&\cite{bc01}&\\
\texttt{BENNER74}&\cite{bcg94}&\cite{vea95}\\
\texttt{BLOSUM62}&\cite{hh92}&\cite{vea95}\\
\texttt{FENG}&\cite{fjd85}&\cite{vea95}\\
\texttt{GONNET}&\cite{gcb92}&\cite{vea95}\\
\texttt{MCLACH}&\cite{m71}&\cite{vea95}\\
\texttt{NWSGAPPEP}&\cite{gb86}&\cite{vea95}\\
\texttt{PAM250}&\cite{dso78}&\cite{vea95}\\
\texttt{RAO}&\cite{r87}&\cite{vea95}\\
\texttt{RUSSELL-RH}&\cite{rssbs97}&\cite{bc01}\\
\texttt{VTML160}&\cite{msv02}&\cite{gb02}\\
\hline
\end{tabular}

%% file: tab3.tex
\begin{tabular}{lcc}
\hline
&\multicolumn{2}{c}{$\varphi_S^{h,g}(\mathcal{A}^r)$}\\
\cline{2-3}
$S$&Global alignments&Local alignments\\
\hline
\texttt{BC0030}&$1.5226$&$1.5060$\\
\texttt{BENNER74}&$1.5601$&$1.5348$\\
\texttt{BLOSUM62}&$1.5532$&$\mathbf{1.5542}$\\
\texttt{FENG}&$1.5062$&$1.4950$\\
\texttt{GONNET}&$1.5419$&$1.5373$\\
\texttt{MCLACH}&$1.5415$&$1.5371$\\
\texttt{NWSGAPPEP}&$1.5306$&$1.5181$\\
\texttt{PAM250}&$1.5243$&$1.5064$\\
\texttt{RAO}&$1.4864$&$1.4912$\\
\texttt{RUSSELL-RH}&$\mathbf{1.4508}$&$1.4508$\\
\texttt{VTML160}&$\mathbf{1.5734}$&$\mathbf{1.4296}$\\
\hline
\end{tabular}

%% file: tab4.tex
\begin{tabular}{lll}
\hline
Cover set&$\mathcal{I}$&$\mathcal{S}$\\
\hline
$C_1$&$\{M,I,L,V\}$&$\{P\}$\\
$C_2$&$\{M,I,L,V,A,P\}$&$\{A,G\}$\\
$C_3$&$\{M,I,L,V,F,W\}$&$\{D,E\}$\\
$C_4$&$\{M,I,L,V,A,P,F,W\}$&$\{N,Q\}$\\
$C_5$&$\{D,E,H,R,K\}$&$\{S,T\}$\\
$C_6$&$\{S,T,Q,N\}$&$\{F,W,Y\}$\\ 
$C_7$&$\{S,T,Q,N,D,E\}$&$\{H,K,R\}$\\
$C_8$&$\{Q,N,D,E,H,R,K\}$&$\{I,L,V\}$\\
$C_9$&$\{S,T,Q,N,D,E,H,R,K\}$&$\{C,F,I,L,M,V,W,Y\}$\\
$C_{10}$&$\{Q,N\}$&$\{D,E,H,K,N,Q,R,S,T\}$\\
$C_{11}$&$\{D,E,Q,N\}$&\\
$C_{12}$&$\{H,R,K\}$&\\
$C_{13}$&$\{R,K\}$&\\
$C_{14}$&$\{F,W,Y\}$&\\
$C_{15}$&$\{G,N\}$&\\
$C_{16}$&$\{A,C,G,S\}$&\\
$C_{17}$&$\{S,T\}$&\\
$C_{18}$&$\{D,E\}$&\\
\hline
\end{tabular}

%% file: tab5.tex
\begin{tabular}{lcccc}
\hline
&\multicolumn{4}{c}{Final values}\\
\cline{2-5}
Parameter&\multicolumn{2}{c}{Global alignments}&\multicolumn{2}{c}{Local alignments}\\
\cline{2-5}
&$\mathcal{I}$ (a)&$\mathcal{S}$ (b)&$\mathcal{I}$ (c)&$\mathcal{S}$ (d)\\
\hline
$b_w$&1&1&1&1\\
$w^-$&0.95&0.85&0.9&0.95\\
$w^+$&2.125&2.25&2.125&2\\
$\lambda$&2&2.875&2&4.375\\
$b_S$&1&1&1&2\\
$S^-$&0.6&1&0.5&1\\
$S^+$&23&10&22.75&10\\
$\mu$&3.125&1&3.625&1.125\\
$h$&29&16.5&29.5&16\\
$g$&0.25&0.75&0.375&0.5\\
\hline
\end{tabular}

%% file: tab6.tex
\begin{tabular}{lccccc}
\hline
$S$&$\varphi_S^{h,g}(\mathcal{A}^{r,1})$&$\rho_{S,1}^{h,g}(\mathcal{A}^{r,1})$&$\rho_{S,2}^{h,g}(\mathcal{A}^{r,1})$&$\rho_{S,3}^{h,g}(\mathcal{A}^{r,1})$&$\rho_{S,4}^{h,g}(\mathcal{A}^{r,1})$\\
\hline
\texttt{BC0030}&$1.5149$&$0.7398$&$0.7738$&$0.7565$&$0.7593$\\
\texttt{BENNER74}&$1.5448$&$0.7607$&$0.7988$&$0.7632$&$0.7661$\\
\texttt{BLOSUM62}&$\mathbf{1.5865}$&$\mathbf{0.7897}$&$\mathbf{0.8278}$&$\mathbf{0.7758}$&$\mathbf{0.7786}$\\
\texttt{FENG}&$1.5216$&$0.7554$&$0.7922$&$0.7457$&$0.7488$\\
\texttt{GONNET}&$1.5253$&$0.7572$&$0.7906$&$0.7494$&$0.7526$\\
\texttt{MCLACH}&$1.5544$&$0.7702$&$0.8045$&$0.7654$&$0.7679$\\
\texttt{NWSGAPPEP}&$1.5019$&$0.7349$&$0.7671$&$0.7489$&$0.7525$\\
\texttt{PAM250}&$1.4996$&$0.7380$&$0.7705$&$0.7436$&$0.7466$\\
\texttt{RAO}&$1.5205$&$0.7553$&$0.7908$&$0.7453$&$0.7486$\\
\texttt{RUSSELL-RH}&$\mathbf{1.4661}$&$\mathbf{0.7274}$&$\mathbf{0.7615}$&$\mathbf{0.7197}$&$\mathbf{0.7227}$\\
\texttt{VTML160}&$1.5789$&$0.7830$&$0.8238$&$0.7736$&$0.7763$\\
\hline
\end{tabular}

%% file: tab7.tex
\begin{tabular}{lc}
\hline
Parameter&Final value\\
\hline
$b_w$&$1$\\
$w^-$&$0.95$\\
$w^+$&$3.625$\\
$\lambda$&$1.875$\\
$b_S$&$1$\\
$S^-$&$1$\\
$S^+$&$5.5$\\
$\mu$&$1.125$\\
$h$&$7.5$\\
$g$&$1.5$\\
\hline
\end{tabular}